\documentclass[onecolumn]{IEEEtran4PSCC}
\ifCLASSINFOpdf
  \usepackage[pdftex]{graphicx}
\else
  \usepackage[dvips]{graphicx}
\fi
\usepackage[cmex10]{amsmath}
\usepackage{amssymb,amsfonts} % \usepackage{amsmath,amssymb,amsfonts}
\usepackage[table]{xcolor}% http://ctan.org/pkg/xcolor
\usepackage{graphicx}  % for pics
\graphicspath{{figs/}}
\usepackage{mathtools}
\usepackage{algorithmic}
\usepackage{textcomp}
\usepackage{url}
\usepackage[mathscr]{euscript}
\usepackage{soul}

\usepackage{multirow}
\newcommand{\tabincell}[2]{\begin{tabular}{@{}#1@{}}#2\end{tabular}}

\usepackage[capitalise]{cleveref}
\crefname{equation}{}{}
\Crefname{equation}{Equation}{Equations}
\crefname{figure}{Fig.}{Figs.}

\usepackage{todonotes}
\presetkeys{todonotes}{inline}{}

\newcommand{\cP}{\boldsymbol{\mathscr{P}}}
\newcommand{\cM}{\boldsymbol{\widehat{\mathscr{P}}}}
\newcommand{\cp}{p}
\newcommand{\cm}{\widehat{p}}

\newcommand{\R}{\mathbb{R}}
\newcommand{\rank}{\operatorname{rank}}

\newcommand{\david}[1]{{\color{black}#1}}

\makeatletter
\let\old@ps@headings\ps@headings
\let\old@ps@IEEEtitlepagestyle\ps@IEEEtitlepagestyle
\def\psccfooter#1{%
    \def\ps@headings{%
        \old@ps@headings%
        \def\@oddfoot{\strut\hfill#1\hfill\strut}%
        \def\@evenfoot{\strut\hfill#1\hfill\strut}%
    }%
    \def\ps@IEEEtitlepagestyle{%
        \old@ps@IEEEtitlepagestyle%
        \def\@oddfoot{\strut\hfill#1\hfill\strut}%
        \def\@evenfoot{\strut\hfill#1\hfill\strut}%
    }%
    \ps@headings%
}
\makeatother

\begin{document}

\title{Baseline Estimation of Commercial Building HVAC Fan Power Using Tensor Completion}

%% To specify the authors when (number of affiliations <= 2)
% \author{
% \IEEEauthorblockN{Shunbo Lei$^1$, David Hong$^2$, Johanna L. Mathieu$^1$, and Ian A. Hiskens$^1$}
% \IEEEauthorblockA{$^1$Department of Electrical Engineering and Computer Science,
% University of Michigan,
% Ann Arbor, MI, USA\\
% $^2$Wharton Statistics Department, University of Pennsylvania,
% Philadelphia, PA, USA\\
% \{shunbol, jlmath, hiskens\}@umich.edu, dahong67@wharton.upenn.edu}
% }% \and
% % \IEEEauthorblockN{Author Name per Affiliation B}
% % \IEEEauthorblockA{(Affiliation B) Department Name of Organization \\
% % Name of the organization, acronyms acceptable\\
% % City, Country\\
% % \{email author n.1, email author n.2\}@domain (if desired)}
% % }

%% To specify the authors when (number of affiliations > 2)
\author{\IEEEauthorblockN{Shunbo Lei\IEEEauthorrefmark{1},
David Hong\IEEEauthorrefmark{2},
Johanna L. Mathieu\IEEEauthorrefmark{1} and
Ian A. Hiskens\IEEEauthorrefmark{1}}
\IEEEauthorblockA{\IEEEauthorrefmark{1} Department of Electrical Engineering and Computer Science\\
University of Michigan,
Ann Arbor, MI, USA\\\{shunbol, jlmath, hiskens\}@umich.edu}
\IEEEauthorblockA{\IEEEauthorrefmark{2} Wharton Statistics Department\\
University of Pennsylvania, Philadelphia, PA, USA\\
dahong67@wharton.upenn.edu}}
% \IEEEauthorblockA{\IEEEauthorrefmark{3} Department Name of Organization C\\
% Name of the organization C,
% Address C\\ Emails if wanted}
% \IEEEauthorblockA{\IEEEauthorrefmark{4}Department Name of Organization D\\
% Name of the organization D,
% Address D\\ Emails if wanted}
% }

\maketitle

\begin{abstract}
  Commercial building heating, ventilation, and air conditioning (HVAC) systems have
  been studied for providing ancillary services to power grids via demand response (DR).
  One critical issue is to estimate the counterfactual baseline power consumption
  that would have prevailed without DR. Baseline methods have been developed based
  on whole building electric load profiles.
  New methods are necessary {\color{black}to estimate the baseline power consumption of HVAC sub-components (e.g., supply and return fans),}
  which have different characteristics compared to that of the whole building.
  Tensor completion can estimate the unobserved entries of multi-dimensional tensors describing complex data sets. It exploits high-dimensional data to capture granular insights  into the problem.
  This paper proposes to use it for baselining HVAC fan power,
  by utilizing its capability of capturing dominant fan power patterns.
  The tensor completion method is evaluated using HVAC fan power
  data from several buildings at the University of Michigan,
  and compared with several existing methods. The tensor completion method generally outperforms the benchmarks.
\end{abstract}


% Use this to place sponsorships
\thanksto{This work was supported in part by the U.S. Department of Energy~{\color{black}via~the project IDREEM: Impact of Demand Response on short and long term~building Energy Efficiency Metrics funded by the Building Technologies Office} under contract number DE-AC02-76SF00515. \david{Hong was supported in part by the NSF BIGDATA grant IIS 1837992 and the Dean's Fund for Postdoctoral Research of the Wharton School.}}

\section{Introduction}

Demand response (DR) is a strategy that incentivizes changes in building electricity consumption to improve grid reliability and economics.
Owing to the large thermal inertia of commercial buildings, their heating, ventilation, and air conditioning (HVAC) systems
can provide ancillary services to electric power systems {\color{black}through} DR while respecting the occupants' comfort.
For example, in \cite{Yashen2014}, the
power consumption of HVAC system fans is controlled to track a regulation signal.

A critical challenge in implementing DR is to estimate the power consumption that would have prevailed without the DR event, referred to as the \textit{baseline}, which is needed for financial settlement {\color{black} and DR impact analysis}.
Therefore, a variety of baseline estimation methods have been developed, generally based on whole building electric load profiles.
Now,
with {\color{black}increasing availability of submetering} in commercial buildings,
it is anticipated that those {\color{black}more granular} data can be utilized to attain more accurate baseline {\color{black}estimates} and more granular
insights into
the {\color{black}responsive} components, e.g., fans in the HVAC systems.
To this end,
new baseline methods are needed,
as the existing methods {\color{black}do not always work well with the new types of data.}

Baseline methods in the literature can be classified into three categories, i.e., averaging,
regression, and control group methods.
They are generally used to baseline whole building electric load.
Averaging methods use the average load~of selected recent days without DR events to estimate the baseline.
They are easy to implement, but typically have large errors~\cite{FeiWANG2018}.
Regression methods fit a linear or non-linear function to~describe the relationship between the load and explanatory
variables such as outdoor temperature, and then use it
to estimate the baseline.
However, their applicability to baseline HVAC fan power is limited,
e.g., due to the weak correlation between HVAC fan power and outdoor temperature~\cite{SLei2019}.
Control group methods identify and utilize the data of similar customers to estimate the baseline.
Normally, a large data set is required~\cite{ZhangYi2016}.

In~\cite{Yashen2014}, signal bandwidth separation is used to estimate the baseline fan power, since the baseline fan power has a much lower bandwidth than
the DR signal (a frequency regulation signal).  However, this approach is only applicable when the DR  signal
is of a high frequency.
In~\cite{Beil2015}~\cite{Keskar2018},
a linear interpolation method is used to baseline fan power.
It uses least squares to fit a linear baseline to the fan power data
{\color{black}over the 5-minute period just before the DR event
and
the 5-minute period after a settling window (i.e., the period during which the fans settle back to their normal operation).}
However, it~does not perform consistently well, generating large errors in some cases.

This paper proposes an approach based on tensor decomposition.
Tensor decomposition is an unsupervised data analysis method that can find
dominant patterns across multiple dimensions,
e.g., time, fan, and day, in the case of submetered fan power data.
See \cite{kolda:2009:tda} for a survey.
It has numerous applications,
from cybersecurity~\cite{Maruhashi2011}
to energy breakdown~\cite{Batra2018}~{\color{black}to}~power~system model reduction~\cite{KaiSUN2019},
and was specifically applied~to~submetered fan power data in {\color{black}our preliminary work}~\cite{hong:2019:eot},
where it captured dominant load patterns in the HVAC supply and return fans.

Tensor completion is the closely related problem of
imputing missing or unobserved entries of a tensor.
% typically accomplished by assuming the unobserved entries
% should generally follow the patterns of the observed entries.
% and it is often accomplished by exploiting within-tensor correlations
% that are estimated from its observed entries,
% e.g., via tensor decomposition.
The ubiquity of tensor structure makes it a fundamental problem
with broad interest and highly active research;
see \cite{song2019tca} for a recent survey.
This paper considers tensor completion via
generalized canonical polyadic (GCP) tensor decomposition \cite{hong:2018:gcp}.
GCP tensor decomposition allows for fitting only observed tensor entries,
making it suitable for tensor completion,
and it has the added ability to do outlier-robust fitting
via an appropriate choice~of fitting loss function.
Other tensor completion approaches could be of interest,
though we leave further exploration for future work.

% Motivated by these findings, we consider a tensor completion approach to baseline estimation.
% Namely, we propose to estimate baseline measurements during DR events
% from a low-rank tensor approximation of measurements outside the DR events.
% The idea is to impute counterfactual baseline measurements
% by encouraging them to agree with correlations across
% time, fan, and date that are found outside the DR events.

This work proposes using GCP tensor completion to estimate HVAC fan power baselines
by considering fan power data within the DR event windows to be missing.
Specifically, we estimate baseline fan power during DR events
from a low-rank GCP tensor approximation of fan power measurements outside the DR events.
Doing so encourages baseline estimates
% to agree with correlations across
% time, fan, and date that occur outside the DR events.
to have load patterns that are consistent
among different fans and different days.
% over different baseline days without DR events.
The tensor completion method is evaluated with power consumption
data of HVAC system supply fans and return fans
that are submetered
in three buildings
at the~University of Michigan,
i.e., Bob and Betty Beyster~Building (BBB), Rackham Graduate School (RAC) and Weill Hall (WH).

The contributions of this work are threefold:
1)~A new method for baselining HVAC fan power is developed based on tensor completion;
2)~The impact of data granularity (i.e., time intervals, aggregation levels) on the method is  investigated; and
3)~The tensor completion method is compared with the three best methods among the ones tested in our previous work~\cite{SLei2019}
using real-world data. Section~\ref{pro_state} provides a brief problem statement.
Section~\ref{introduce_method} introduces the tensor completion method.
Section~\ref{case_studies} presents case studies.
Section~\ref{conclusion_future_work} concludes the paper and discusses the future work.

\section{Problem Statement}\label{pro_state}

\begin{figure}[b!] % [t!] % [htbp]
  \centering
  \includegraphics[width=3in]{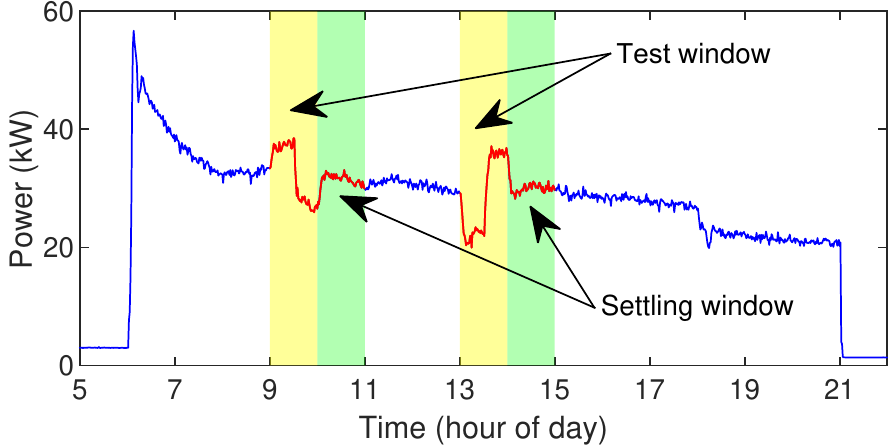}
  \caption{Total fan power trace of RAC on an event day (Sep. 26, 2017).}
  \label{RAC_example}
%\vspace{-1mm}
\end{figure}

\begin{figure}[b!] \centering
% \begin{subfigure}{.5\linewidth} \centering
\includegraphics[scale=0.91]{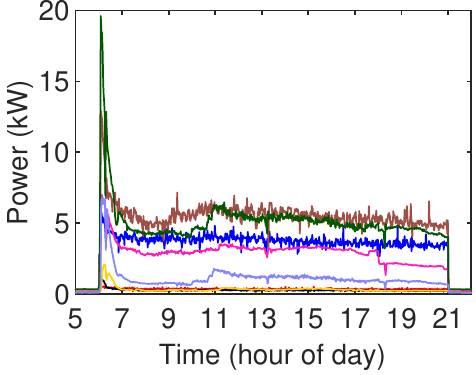}%\caption{...}
% \label{fig:pow:traces:20170905}
% \end{subfigure}%
% \begin{subfigure}{.5\linewidth} \centering
\includegraphics[scale=0.91]{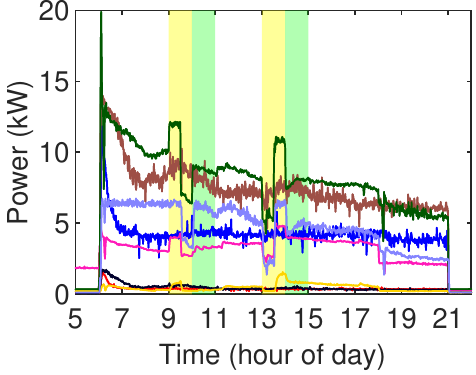}%\caption{...}
% \label{fig:pow:traces:20170926}
% \end{subfigure}
\\
\includegraphics[scale=0.74]{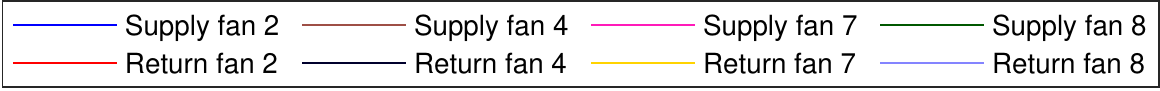}
%\vspace{-1mm}
\caption{Per-fan power traces of RAC on a baseline day (Left, Sep. 5, 2017)
and an event day (Right, Sep. 26, 2017). One-minute interval data are shown.}
\label{fig:pow:traces}
% \vspace{-5mm}
%\vspace{-1mm}
\end{figure}

Fig.~\ref{RAC_example} shows two DR tests on RAC\@.
The morning event is called an up-down test.
During the test window (9-10am), room temperature
setpoints are decreased below nominal values for 30 minutes
and then increased symmetrically above nominal values for 30 minutes,
causing HVAC fan power to go up and then go down.
The fans return to their normal operation after a settling window, which we assume to be one hour.
The afternoon event is a down-up test with opposite
room temperature setpoint changes.
These tests are designed to assess the energy efficiency of fast
DR actions, which primarily impact the~fans~{\color{black}\cite{Keskar2019}}.
To this end, we need to estimate the fan power that would have
prevailed without DR events, i.e., the
baseline fan power during each event window (including the test window and
settling window). In our case study we differentiate between the morning event window (9-11am) and the afternoon event window
(1-3pm).

A commercial building HVAC system normally has multiple supply fans and return
fans. Our submetering
equipment provides us with
1-minute resolution power data from each~fan.
That is, while our task is baselining the total fan power, we have
data of a higher granularity, i.e., the power {\color{black}consumption} of each~fan.
Fig.~\ref{fig:pow:traces} (Right) shows the power trace of each fan in RAC on the same test day as used in Fig.~\ref{RAC_example},
while Fig.~\ref{fig:pow:traces} (Left) shows fan power traces  on
a day without DR events, i.e., a \textit{baseline day}.
% {\color{blue}Data of baseline days indicate the normal operation of HVAC fans without
% DR events, thus are informative in estimating the baseline fan power on DR event days,}
{\color{black}Data \david{from} baseline days {\color{black}corresponds to} normal operation of HVAC fans without
DR events, \david{and thus can be used to estimate} the
baseline fan power on DR event days.}

Specifically, our baseline estimation problem can be described as follows:
{\emph{Given the power profile of each fan in time slots outside event window(s)
on days with DR event(s) and the power profile of each fan on baseline days,
estimate the total fan power that would have prevailed in
time slots within event window(s)
on event days if DR event(s) did not happen.}} Other baseline methods may have different data
requirements; the above problem description is for our tensor completion method.

\begin{table}[t!]
  \centering
  \caption{Summary of Data Sets}
  %\vspace{-1mm}
  \setlength{\tabcolsep}{5pt}
  \begin{tabular}{ccccc}
    \hline
    \multirow{3}[0]{*}{} &
    \multirow{3}[0]{*}{\tabincell{c}{\# of fans\\(SF: supply fan; \\RF: return fan.)}} & \multirow{3}[0]{*}{\tabincell{c}{\# of \\baseline days}} &
    \multicolumn{2}{c}{\multirow{2}[0]{*}{\tabincell{c}{Total fan power \\in day mode (kW)}}} \\
    & &  & \multicolumn{2}{c}{} \\
    \cline{4-5}
    & & & Peak & Average\\
    \hline
    BBB-2017 & 1 SF, 1 RF & 55 (in Jun.-Oct.) & \;\;\;35.8\;\;\; & 12.2 \\
    %\hline
    BBB-2018 & 4 SFs, 3 RFs & 16 (in Oct.) & 105.3 & 38.3 \\
    %\hline
    RAC-2017 & 4 SFs, 4 RFs & 49 (in Jul.-Oct.) & 63.6 & 18.7 \\
    %\hline
    RAC-2018 & 4 SFs, 4 RFs & 30 (in May-Oct.) & 63.0 & 24.7 \\
    %\hline
    WH-2017 & 2 SFs, 2 RFs & 86 (in Jun.-Oct.) & 125.9 & 45.7\\
    \hline
\end{tabular}%
\label{data_table}%
%\vspace{-5mm}
\end{table}%

We evaluate the accuracy of the proposed baseline estimation method using data from baseline days
{\color{black}in a data set}.
As no DR events happen on {\color{black}a baseline day},
the measured total fan power is
the true baseline during event windows (without tests) {\color{black}for the same baseline day}.
The errors~of~the~proposed~baseline estimation method can be
assessed by comparing the estimates with the measurements.
Table~\ref{data_table} lists the five data sets used in this work, corresponding to five different building-years.
 To estimate the baseline in event window(s) on a given day~of~a building-year,
only data from the same building-year is used.

\section{Proposed Tensor Completion Method}\label{introduce_method}
% !TEX root = main.tex

\begin{figure}[t!] % [t!] % [htbp]
  \centering
  \includegraphics[width=3.45in]{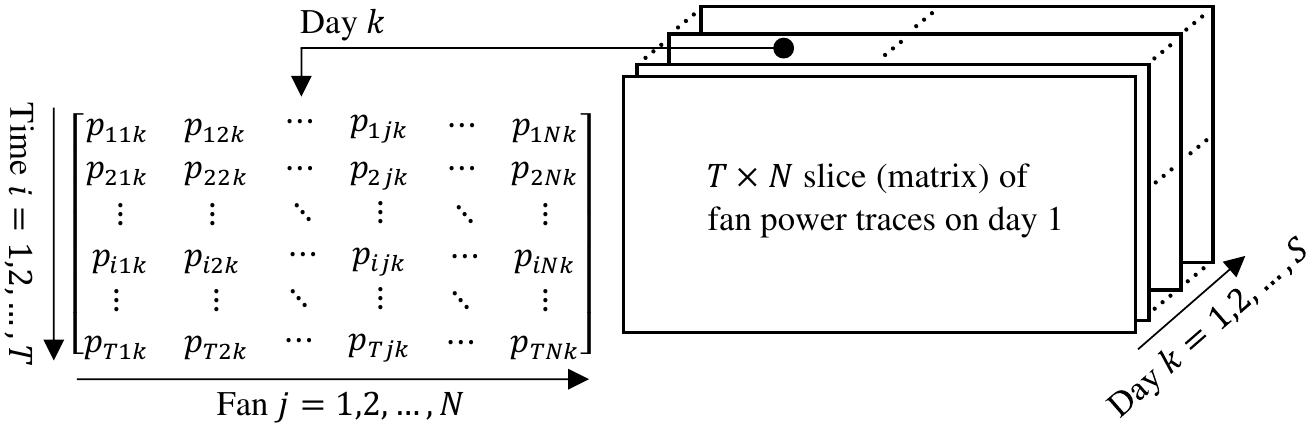}
  \caption{Formation of a three-way tensor $\cP$ based on HVAC fan power data.}
  \label{fig:tensor:example}
\end{figure}

Motivated by \cite{hong:2019:eot},
which uses the canonical polyadic (CP) tensor decomposition
to capture dominant fan power behavior,
we investigate using tensor completion
to form baseline estimates.
As shown in \cref{fig:tensor:example},
we form the fan power data into a
\emph{three-way} time $\times$ fan $\times$ day tensor\footnote{{\color{black}The fans and days can be arbitrarily ordered in the tensor, which does not impact the results of the  proposed baseline method in this work.}},
i.e., a three-dimensional $T \times N \times S$ array
$\cP \in \R^{T \times N \times S}$
whose $(i,j,k)$th entry $\cp_{ijk}$ is the power
at time slot $i$ of fan $j$ on day $k$,
with $i=1,2,...,T$, $j=1,2,...,N$ and $k=1,2,...,S$.
Using this representation,
correlations across fans and days
become naturally expressed as patterns
across each of the three \emph{modes} (time, fan, and day),
and can be captured by tensor decomposition.
%See \cite{kolda:2009:tda} for a survey of tensor decomposition.

\begin{figure}[t!] % [t!] % [htbp]
  \centering
  \includegraphics[width=3.45in]{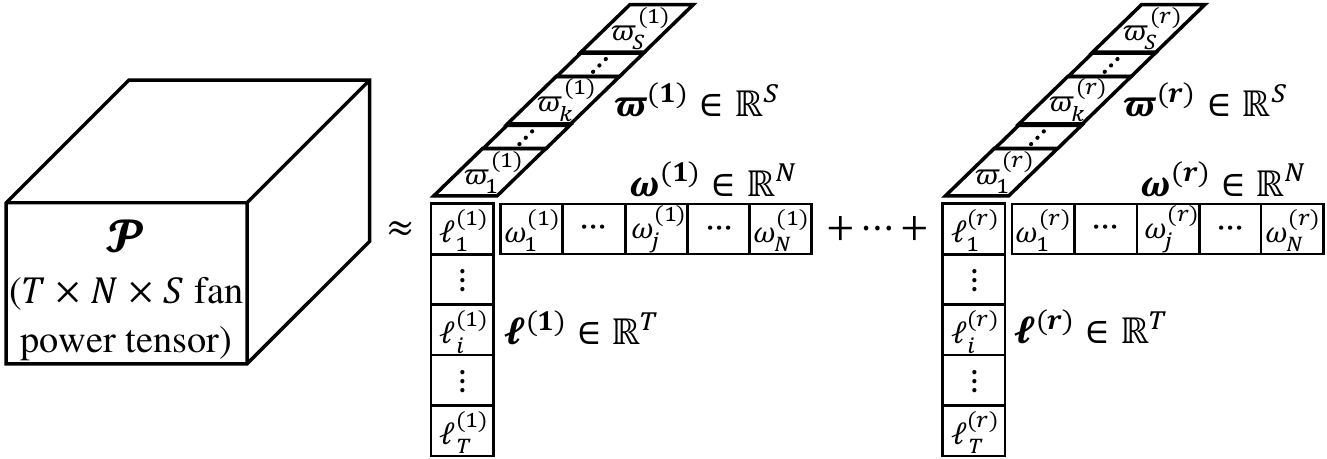}
  \caption{Rank-$r$ %canonical polyadic (CP)
  tensor decomposition of the fan power tensor.
  The left-hand side of the approximation is the fan power tensor $\cP$.
  The right-hand side is a sum of $r$ outer products, referred to as $\cM$.
  (Time slot index $i=1,2,...,T$; fan index $j=1,2,...,N$; and day index $k=1,2,...,S$.)}
  \label{tensor_decom}
% \vspace{-4mm}
\end{figure}

In the proposed method, we include data from $S-1$ baseline days
and $1$ event day in $\cP$.
We consider the to-be-estimated baseline power within DR event window(s) of the event day
as missing or unobserved measurements,
and impute these tensor entries
by approximating the remaining (known) entries with a low-rank tensor $\cM$
as shown in \cref{tensor_decom}.
Namely, we estimate the counterfactual measurements by the entries of $\cM$,
where $\cM$ is rank (at most) $r$.
As illustrated in Fig.~\ref{tensor_decom}, $\cM$ is a sum of $r$ outer products:
\begin{equation} \label{eq:sumouterproducts}
  \cM =
  \ell^{(1)} \circ \omega^{(1)} \circ \bar{\omega}^{(1)}
  + \cdots +
  \ell^{(r)} \circ \omega^{(r)} \circ \bar{\omega}^{(r)}
.
\end{equation}
% {\color{blue}In \eqref{eq:sumouterproducts}, $\circ$ denotes an outer product, as shown in Fig.~\ref{tensor_decom}.~Expressed in terms of its entries,
% \eqref{eq:sumouterproducts} is equivalent to:
{\color{black}\david{where} $\circ$ denotes an outer product,
as shown in \cref{tensor_decom}.
\david{Namely, the entries of $\cM$ are}:
\begin{equation} \label{eq:componentwiseexpression}
  \cm_{ijk} =
  \ell^{(1)}_i \omega^{(1)}_j \bar{\omega}^{(1)}_k
  + \cdots +
  \ell^{(r)}_i \omega^{(r)}_j \bar{\omega}^{(r)}_k
.
\end{equation}}The $(i,j,k)$th entry $\cm_{ijk}$ of $\cM$ is the approximated power
at time slot $i$ of fan $j$ on day $k$.
The vectors forming $\cM$ capture dominant underlying patterns across each mode:
$\ell^{(1)},...,\ell^{(r)} \in \R^T$ are factors along the time mode,
$\omega^{(1)},...,\omega^{(r)} \in \R^N$ are factors along the fan mode,
and
$\bar{\omega}^{(1)},...,\bar{\omega}^{(r)} \in \R^S$ are factors along the day mode.

\begin{figure}[t!] % [t!] % [htbp]
  \centering
  \includegraphics[width=2in]{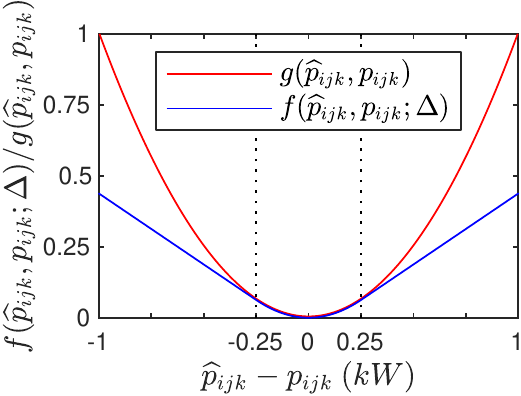}
  \caption{Huber loss function ($f$) with $\Delta=0.25$ and $L_2$-norm loss function~($g$).}
  \label{huber_cost}
% \vspace{-4mm}
\end{figure}

We obtain $\cM$ via the GCP tensor decomposition \cite{hong:2018:gcp}
of $\cP$ that minimizes the following loss function:
\begin{equation} \label{eq:gcp:objective}
  F(\cM,\cP)
  \coloneqq
  \sum_{(i,j,k) \in \boldsymbol{\Omega}} f(\cm_{ijk},\cp_{ijk})
  ,
  \quad
  \text{s.t. }
  \rank \cM \leq r
  .
\end{equation}
where the sum is over indices $(i,j,k) \in \boldsymbol{\Omega}$ which are known, i.e., not to be estimated.
{\color{black}Namely, indices $(i,j,k)$ in the set $\boldsymbol{\Omega}$ % correspond to
% all entries associated with baseline days and the entries associated with the event day but outside the DR event window(s).}
\david{include:
a) all entries from baseline days, and
b) all entries outside the DR event window(s) from event days.}}
% In particular, a
\david{A} commonly-used loss function is the $L_2$-norm loss function, i.e., the squared error loss function. For a given $(i,j,k)\in \boldsymbol{\Omega}$,
this loss is calculated by:

\begin{equation}
\begin{aligned}
g(\cm_{ijk},\cp_{ijk})\coloneqq\left(\cm_{ijk}-\cp_{ijk}\right)^2.
\end{aligned}
\end{equation}
In this work, we use the Huber loss function instead \cite{huber1964reo}:
\begin{equation}
  f(\cm_{ijk},\cp_{ijk}; \Delta)
  \coloneqq
  \begin{cases}
    g(\cm_{ijk},\cp_{ijk}),
    \,\,\text{if } | \cm_{ijk}-\cp_{ijk} | \leq \Delta;
    \\
    2\Delta | \cm_{ijk}-\cp_{ijk} | - \Delta^2,
    \,\,\,\,\,\text{otherwise}.
  \end{cases}
\end{equation}
\Cref{huber_cost} compares this loss function ($\Delta=0.25$)
with the usual squared error loss of conventional CP tensor decomposition.
The Huber loss penalizes the square of small residuals between $\cM$ and $\cP$,
encouraging a conventional least-squares fit of $\cM$ to $\cP$,
but penalizes large residuals, i.e., potential outliers, by only their magnitudes.
Doing so can make the results more robust to outliers in the data \cite{hastie2009teo}.
In our experiments, we find that
using the Huber loss function attains much smaller baseline errors
than using the squared error function for data sets BBB-2018 and WH-2017,
which have easily identifiable outliers.

To summarize, we propose to:
1) \emph{approximate}%
   \footnote{\emph{All} entries of $\cP$ are approximated in $\cM$,
     including those on baseline days and in the event day but outside the DR event window(s), even though they are known and do not need to be estimated.}
   the known measurements in the fan power tensor $\cP$
   with a rank-$r$ tensor $\cM$ by GCP tensor decomposition;
   then
2) \emph{estimate} counterfactual measurements
    using the entries of $\cM$.
This approach estimates baselines by encouraging a low-rank tensor structure
and is a tensor generalization of the low-rank matrix completion.
Tensor decomposition serves as a tensor analogue
to Principal Component Analysis (PCA),
where the factors in \cref{eq:sumouterproducts} correspond to principal components.
The generalization enables this tensor approach to exploit correlation
along time, fan, and day.

\subsection{Optimization Algorithm for GCP Tensor Decomposition}
% !TEX root = main.tex

We use the implementation of GCP tensor decomposition
in the MATLAB Tensor Toolbox \cite{TTB_Software}
that minimizes \cref{eq:gcp:objective}
with respect to the factors
$\ell^{(1)},...,\ell^{(r)} \in \R^T$,
$\omega^{(1)},...,\omega^{(r)} \in \R^N$,
and
$\bar{\omega}^{(1)},...,\bar{\omega}^{(r)} \in \R^S$
of $\cM$ as defined in \cref{eq:sumouterproducts}.
The toolbox function \texttt{gcp\_opt} minimizes
this non-convex optimization problem
via a first-order algorithm, i.e., the limited-memory BFGS with bound constraints (L-BFGS-B) algorithm \cite{byrd1995alm},
using the special form of the gradient \cite[Section 4.1]{hong:2018:gcp}.
Further details are available in~\cite{hong:2018:gcp}.
This approach is not guaranteed to always find a global minimum,
so we run four trials with different initializations and select the best trial,
i.e., the trial with the smallest final objective function value.

% For each tensor completion problem,
% we run 4 trials with different random initializations.
% The solution with the smallest objective function value
% is~selected.

\subsection{Selecting a Rank} \label{sec:proposal:rank}
% !TEX root = main.tex

The proposed approach requires that we choose the
approximation rank $r$.
Using a rank that is too large can underconstrain $\cM$.
For example, for any $\cM \in \R^{T \times N \times S}$ \cite{kruskal1989rda}:
\begin{equation}
  \rank \cM \leq \min(TN,TS,NS)
  .
\end{equation}
Thus, for $r = \min(TN,TS,NS)$,
any tensor $\cM$ matching~$\cP$ on the known entries
minimizes \cref{eq:gcp:objective}
regardless of how it estimates counterfactual baselines,
making the estimates arbitrary.
On the other hand, a rank that is
too small can insufficiently capture
the diversity of fan power behavior,
forming baseline estimates from
only broad coarse-grain fan power patterns.
Hence, there is a trade-off between
data overfit that is incurred at large ranks
and approximation error that is incurred at small ranks.
This trade-off is affected both
by how low-rank baseline fan power behavior is,
which is unknown,
and by how much of the tensor needs to be estimated,
making optimal rank selection challenging.

Fitting higher rank approximations to the data can also take more time,
and in our preliminary experiments,
a rank of 12 seemed to reasonably balance these various trade-offs.
Using higher rank approximations increased computation time
without yielding large performance improvements.
Thus, we use a rank of 12 for the tensor completions in this paper.
Developing systematic approaches to rank selection
for baseline estimation is important future work.

% In our preliminary experiments, a rank higher than 12 greatly
% increases the computation time of our tensor completion problems, and only leads to small performance improvements.
% Thus, rank-12 tensor completions are conducted.
% For each tensor completion problem,
% we run 4 trials with different random initializations.
% The solution with the smallest objective function value
% is~selected.

% This paper selects the rank
% based on our preliminary experiments.

% \todo{discuss typical rank? [David]}
% =======
% that seem to perform best in our experiments;
% developing systematic approaches to rank selection
% for baseline estimation is an important future work.
% >>>>>>> 7d085afd018c8a44f691c9918600f9c18fe9d4ea

\subsection{Data Granularity}

As mentioned,
the data we obtain from the submeters is 1-minute resolution power
data
for each HVAC fan.
As~reported in~\cite{Jain2014}~\cite{Raffi2018},
the performance of an electric load forecasting method can be impacted by the
data temporal and spatial granularity.
Therefore, we  investigate the impact of data granularity on the performance of our method.
We test our method using 1-minute, 5-minute, 15-minute and 30-minute interval data.
{\color{black}We select 5-minute, 15-minute and 30-minute intervals because these are frequently used in the electric load forecasting and prediction literature}
{\color{black}\cite{ZhangYi2016}
\cite{Raffi2018}-\cite{Sun2019}.
We do not test the method using data of longer intervals,}
{\color{black}as we are interested in baselining fan power at subhourly timescales corresponding to our DR experiments.} %\cite{Coughlin2009} \cite{Guan2013} \cite{Sun2019} \cite{Raffi2018}
In this work, the 5-minute data are obtained by averaging the original 1-minute data for every 5 minutes, and
likewise for the 15-minute and 30-minute data sets.
{\color{black}In \cite{hong:2019:eot}, HVAC fan power patterns are investigated with
per-fan power data and total fan power data, respectively.}
We also test our method using per-fan power data versus total fan power data in this work.
In the latter case, although the fan power tensor's dimension becomes $T\times S$,
the tensor completion method is still applicable.

\subsection{Performance Metrics}

Three metrics are used to evaluate the proposed baseline method's performance.
%Let $\cp_{ik}=\sum_j \cp_{ijk}$ and $\cm_{ik}=\sum_j \cm_{ijk}$ be the measured and estimated total fan
Let $\cp_{i}$ and $\cm_{i}$ be the measured and estimated total fan
power at time slot~$i$, respectively.
Also let $\boldsymbol{\tau}=\{\tau_s,\tau_{s+1},...,\tau_e\}$
be the time slots within an event window.
The first metric is the coefficient of variation (CV),
{\color{black}which is also referred to as the
coefficient of~variance~of~root mean squared error (CV-RMSE).}
It measures the baseline accuracy using the ratio of
the standard deviation of~estimation~errors~to the mean of the true
values. For each event window
CV can~be calculated by~\cite{ASHRAE_14_2014}:
\begin{equation}\label{CV_eq}
  CV (\%) = \frac{\sqrt{\frac{1}{\left| \boldsymbol{\tau} \right| -1} \sum_{i=\tau_s}^{\tau_e} \left(\cm_i - \cp_i\right)^2}}
  {\frac{1}{\left| \boldsymbol{\tau} \right|}\sum_{i=\tau_s}^{\tau_e}\cp_i} \times 100.
\end{equation}
% {\color{blue}As indicated in \eqref{CV_eq}, CV is actually the RMSE nomarlized by
% the mean of the true values.}
{\color{black}\david{Namely}, CV is \david{the} RMSE \david{normalized} by
the \david{true mean value.}}

The second metric is the
normalized mean bias error (NMBE)\@. Positive and negative values indicate
overestimation and underestimation of the baseline, respectively.
For each event window
NMBE can be calculated by~\cite{ASHRAE_14_2014}:
\begin{equation}
  NMBE (\%) = \frac{\frac{1}{\left| \boldsymbol{\tau} \right| -1} \sum_{i=\tau_s}^{\tau_e} \left(\cm_i - \cp_i\right)}
  {\frac{1}{\left| \boldsymbol{\tau} \right|}\sum_{i=\tau_s}^{\tau_e}\cp_i} \times 100. \label{NMBE_eq}
\end{equation}
% {\color{blue}Note that the CV and NMBE metrics are
{\color{black} \david{CV and NMBE are}
recommended~by the American Society of Heating, Refrigeration and Air Conditioning Engineers (ASHRAE)
and \david{are} commonly adopted~in~electric load forecasting or prediction studies \cite{Jain2014}~\cite{Raffi2018}~\cite{Edwards2012}.
ASHRAE also prescribes the acceptable tolerances for those two metrics \cite{ASHRAE_14_2014}.}

The third metric is the additional energy consumption (AEC),
which compares the energy consumption during an event window of an event day with the baseline~\cite{Keskar2018}.
While it was designed to assess energy efficiency,
it can also be used to assess baseline method performance
because the AEC will be zero if the baseline method is perfectly {\color{black}unbiased}.
Let $\delta$ indicate the data temporal resolution in minutes. For example, for 15-minute interval data, $\delta=15$.
For each event window
AEC can be calculated by:
\begin{equation}
  AEC (\mathrm{kWh}) = \sum_{i=\tau_s}^{\tau_e} \left(\cm_i - \cp_i\right) \times \frac{\delta}{60}. \label{AEC_eq}
\end{equation}
Note that AEC is similar to NMBE without normalization.
{\color{black}It is a metric to measure the bias of baseline methods.}
It is used here to provide more straightforward assessment of baseline errors in terms of energy consumption.

For each data set, leave-one-out cross-validation {\color{black}\cite{Mathieu2011_1}} is conducted to attain metric statistics. Specifically, the baseline fan power within event windows on a given baseline day $k$ is assumed unknown and estimated based on all other entries of
$\cP$. The estimate is compared with the true baseline to calculate metric values. This process is repeated for each baseline day $k=1,2,...,S$. Metric statistics thus are obtained based on the values of each metric for each event window on different~days.

\section{Case Studies}\label{case_studies}

%This section applies the proposed baseline method to the five data sets in Table~\ref{data_table}.
% The MATLAB Tensor Toolbox~\cite{TTB_Software}
% is used to process tensor data and
% solve tensor completion optimization problems
% via L-BFGS-B \cite{byrd1995alm}.

% {\color{blue}In this section, we apply the proposed baseline~method~and benchmark methods to the data sets listed in Table~\ref{data_table}.~As~mentioned in Section~\ref{pro_state}, the baseline methods are evaluated using data from baseline days. As no DR events
{\color{black}In this section, we apply the proposed baseline~method~and benchmark methods to the data sets listed in \cref{data_table}. As \david{described} in \cref{pro_state}, the baseline methods are evaluated using data from baseline days. \david{Since} no DR events
happen on baseline days, the measured actual total fan power is the true baseline. {\color{black}Errors of
a baseline method can be assessed by
comparing~its baseline estimates with the true baselines (i.e., the measurements).}}
% As mentioned, for each data set, leave-one-out cross-validation is conducted to attain metric statistics.}
% \david{For} each data set, leave-one-out cross-validation is conducted to attain metric statistics.}

\subsection{Impact of Temporal Frequency of the Data}

\newcommand{\sval}[1]{\textbf{#1}}        % smallest value
\newcommand{\lval}[1]{#1} % largest value
\begin{table}[b!]
  \centering
  \caption{Mean values and Standard Deviations ($\sigma$) of CV and NMBE for Different Temporal Frequencies of the Data}
  \setlength{\tabcolsep}{1.5pt}
  \begin{tabular}{c|c|cccc}
    \hline
    \multicolumn{2}{c}{}  & 1-min & 5-min & 15-min & 30-min\\
    \hline
    \multirow{5}[0]{*}{\rotatebox{90}{\tabincell{c}{CV (\%), \\9-11am: \\mean ($\sigma$)}}}
    & BBB-2017  & \lval{15.63} (1.93)  & 8.16 (1.97)   &  \sval{6.12 (1.79)} & 6.12 (2.49)   \\
    %\cline{2-6}
    & BBB-2018 & \lval{10.01} (2.21)  &  \sval{5.23 (1.71)} & 5.25 (3.49)  & 5.32 (4.02)  \\
    %\cline{2-6}
    & RAC-2017 & 10.20 (7.26) &  \sval{8.75 (5.99)} & 9.14 (6.77)  & \lval{10.27 (8.77)}  \\
    %\cline{2-6}
    & RAC-2018 & 8.15 (9.51) & 7.98 (8.34) &  \sval{7.06 (7.89)} & \lval{8.76 (9.94)}  \\
    %\cline{2-6}
    & WH-2017 & \lval{10.71} (6.05) & 10.39 (4.35) &  \sval{10.19} (5.28) & 10.51 (6.20) \\
    \hline
    \multirow{5}[0]{*}{\rotatebox{90}{\tabincell{c}{CV (\%), \\1-3pm: \\mean ($\sigma$)}}}
    & BBB-2017  & \lval{13.31} (2.96) & 7.49 (3.26) &  \sval{6.22 (2.95)} & 6.52 (4.14) \\
    %\cline{2-6}
    & BBB-2018 & \lval{9.72 (4.01)}  & 6.50 (3.03) &  \sval{6.42 (2.70)}  & 6.58 (3.40)  \\
    %\cline{2-6}
    & RAC-2017 & \lval{4.45 (2.77)} & 3.61 (1.85) &  \sval{3.25 (1.74)}  & 3.28 (2.75)  \\
    %\cline{2-6}
    & RAC-2018 & 5.54 (5.73)  & 5.01 (5.11) &  \sval{4.59 (4.18)}  & \lval{5.63 (6.00)}  \\
    %\cline{2-6}
    & WH-2017 & \lval{12.01 (7.62)} & 10.27 (5.58) &  \sval{9.47} (5.90) & 9.72 (6.52) \\
    \hline
    \multirow{5}[0]{*}{\rotatebox{90}{\tabincell{c}{NMBE (\%), \\9-11am: \\mean ($\sigma$)}}}
    & BBB-2017 & \lval{-4.45} (4.60) & -0.23 (4.31) & -0.15 (4.43) &  \sval{-0.05} (6.07) \\
    %\cline{2-6}
    & BBB-2018 & \lval{-5.13} (3.74) &  \sval{-0.37 (3.62)} & -0.95 (5.78)  & 1.08 (6.31)  \\
    %\cline{2-6}
    & RAC-2017 & -1.88 (13.09) &  \sval{-1.69 (10.52)} & \lval{-1.96} (10.62)  & -1.93 (14.32)  \\
    %\cline{2-6}
    & RAC-2018 & 1.13 (10.01) & 1.01 (8.98) &  \sval{0.99 (8.71)} & \lval{1.57 (13.33)}  \\
    %\cline{2-6}
    & WH-2017 & -0.85 (10.50) & -0.56 (9.82) &  \sval{-0.43} (10.04) & \lval{0.94 (12.47)} \\
    \hline
    \multirow{5}[0]{*}{\rotatebox{90}{\tabincell{c}{NMBE (\%), \\1-3pm: \\mean ($\sigma$)}}}
    &BBB-2017 & \lval{-4.52} (5.33) & -0.40 (5.27) &  \sval{-0.29 (5.23)} & -0.89 (7.98) \\
    %\cline{2-6}
    & BBB-2018 & \lval{-4.56} (6.02) &  \sval{0.68} (6.32) & 0.84 (6.53)  & -0.71 (7.83)  \\
    %\cline{2-6}
    & RAC-2017 & 0.32 (3.15) & \lval{-0.37} (2.92) & -0.14 (2.89)  &  \sval{0.09} (4.10)  \\
    %\cline{2-6}
    & RAC-2018 & -0.84 (6.06) & \lval{-1.15} (6.57) & -0.75 (6.30)  &  \sval{-0.72} (9.05)  \\
    %\cline{2-6}
    & WH-2017 &  \sval{-0.38} (10.24) & -0.59 (8.03) & -0.41 (8.65) & \lval{-1.05} (10.12) \\
    \hline
\end{tabular}%
\label{temporal_selection}%
%\vspace{-4mm}
\end{table}%

We first investigate the impact of the data temporal frequency
on the performance of our baseline method.
We apply the method to the original 1-minute interval data,
and also 5-minute, 15-minute and 30-minute interval data.
% In our preliminary experiments, a rank higher than 12 greatly
% increases the computation time of our tensor completion problems, and only leads to small performance improvements.
% Thus, rank-12 tensor completions are conducted.
% For each tensor completion problem,
% we run 4 trials with different random initializations.
% The solution with the smallest objective function value
% is~selected.
Table~\ref{temporal_selection} reports the means and standard
deviations of CV and NMBE for the morning and afternoon event windows
of our five data sets.
The smallest mean value in each row is bolded;
the associated standard deviation is bolded if it is also the smallest.
% In each row, the smallest (resp.~largest) mean value of the metric is
% bolded but not italicized (resp.~bolded and also italicized),
% and the standard deviation is similarly formatted if it is the smallest (resp.~largest).

In~\cite{ASHRAE_14_2014}, the suggested acceptable tolerances
for NMBE and CV are $\pm5\%$ and $15\%$ when using monthly data,
and $\pm10\%$ and $30\%$ when using hourly data.
In this regard, with sub-hourly data here,
the tensor completion method's performance is generally inline with the above suggested tolerances when using hourly data.
Still, higher precision and accuracy of baseline estimates are
needed for DR applications, which involve financial settlement~\cite{FeiWANG2018}.

As indicated in Table~\ref{temporal_selection},
our method with 15-minute interval data generally has the best performance
{\color{black}compared to that with 1-minute, 5-minute or 30-minute interval data.}
It attains the smallest mean value and standard deviation for CV in most cases and
for NMBE in some cases.
(Normally, it is much more difficult to achieve a lower CV than a
lower NMBE according to~\cite{ASHRAE_14_2014}.)
In the other cases, its performance is still comparable to
the best outcome.

\begin{figure}[t!] % [t!] % [htbp]
  \centering
  \includegraphics[width=3.45in]{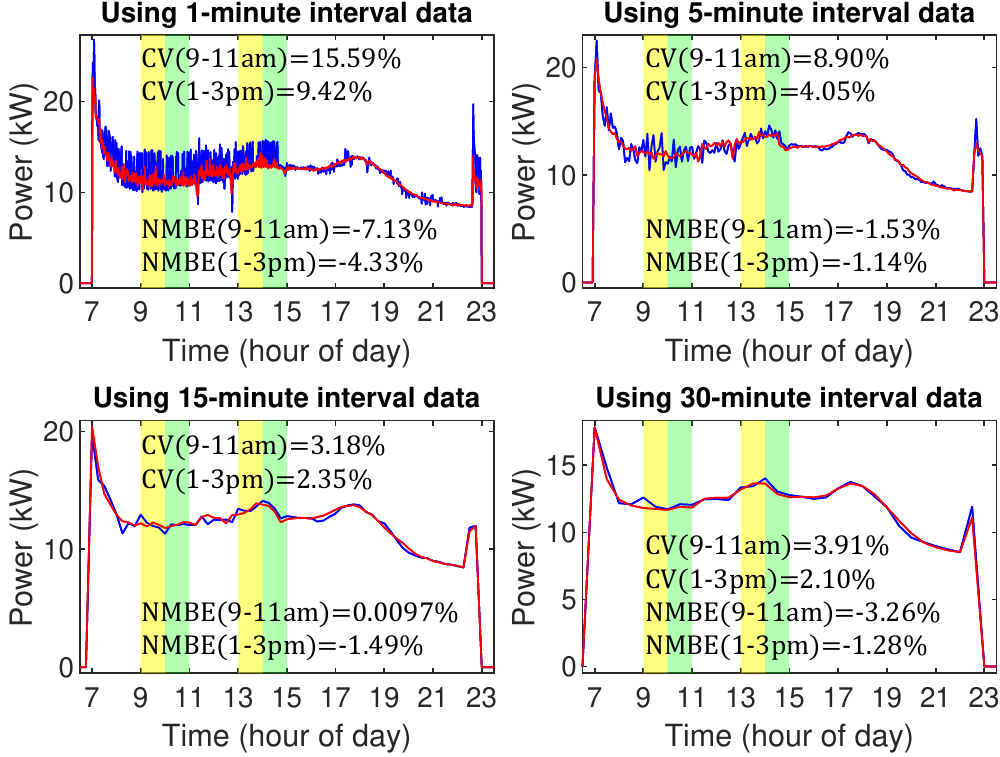}\\
  \includegraphics[width=3.45in]{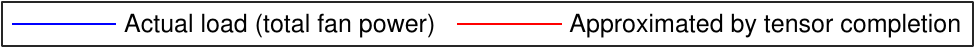}
  \caption{Actual and estimated total fan power traces based on given data with different temporal frequencies (BBB, Aug. 9, 2017).}
  \label{actual_n_approx}
% \vspace{-4mm}
\end{figure}

Fig.~\ref{actual_n_approx}
shows an example of the actual total fan power traces and tensor
completion approximations using data with different temporal intervals.
The tensor completion method cannot effectively capture the high frequency variation seen in the 1-minute interval data, resulting in higher values of CV and NMBE. In most cases, small NMBE values are attained using 5-minute or 15-minute interval data, but the 5-minute interval data yields higher CV values as the data still has high frequency variation.

\subsection{Per-Fan Power Data versus Total Fan Power Data}

\begin{figure} \centering
%   \begin{subfigure}{.2\linewidth} \centering
    \includegraphics[scale=0.62]{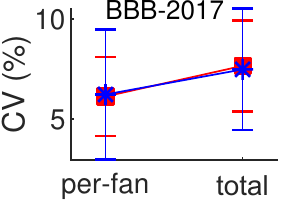}%
%     \label{cv1}
%   \end{subfigure}\hfil
%   \begin{subfigure}{.2\linewidth} \centering
    \includegraphics[scale=0.62]{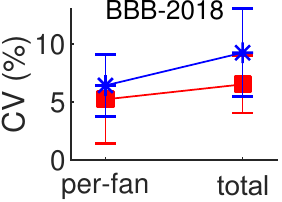}%
%     \label{cv2}
%   \end{subfigure}\hfil
%   \begin{subfigure}{.2\linewidth} \centering
    \includegraphics[scale=0.62]{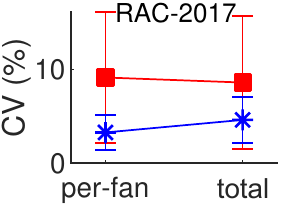}%
%     \label{cv3}
%   \end{subfigure}\hfil
%   \begin{subfigure}{.2\linewidth} \centering
    \includegraphics[scale=0.62]{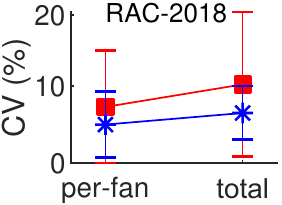}%
%     \label{cv4}
%   \end{subfigure}\hfil
%   \begin{subfigure}{.2\linewidth} \centering
    \includegraphics[scale=0.62]{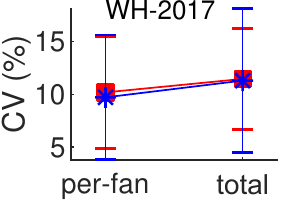}\\[1.5mm]
%     \label{cv5}
%   \end{subfigure}\\
%   \vspace{-2.5mm}
%   \begin{subfigure}{.2\linewidth} \centering
    \includegraphics[scale=0.62]{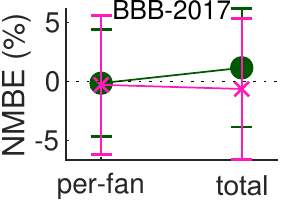}%
%     \label{MBE1}
%   \end{subfigure}\hfil
%   \begin{subfigure}{.2\linewidth} \centering
    \includegraphics[scale=0.62]{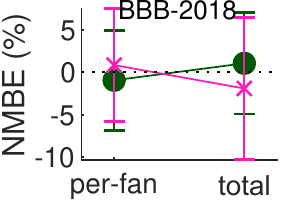}%
%     \label{MBE2}
%   \end{subfigure}\hfil
%   \begin{subfigure}{.2\linewidth} \centering
    \includegraphics[scale=0.62]{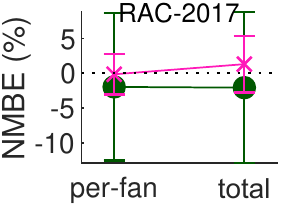}%
%     \label{MBE3}
%   \end{subfigure}\hfil
%   \begin{subfigure}{.2\linewidth} \centering
    \includegraphics[scale=0.62]{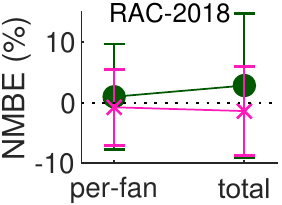}%
%     \label{MBE4}
%   \end{subfigure}\hfil
%   \begin{subfigure}{.2\linewidth} \centering
    \includegraphics[scale=0.62]{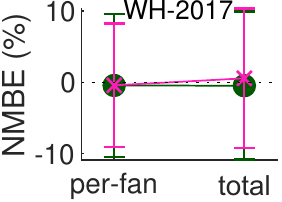}\\
%     \label{MBE5}
%   \end{subfigure}\\
%    \vspace{-4mm}
    \includegraphics[scale=0.74]{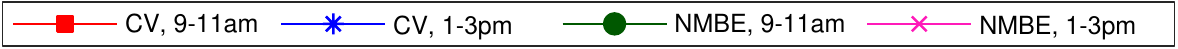}
    \caption{Mean values of CV and NMBE of the tensor completion method with per-fan and total fan power data. (The error bars represent standard deviations.)}
    \label{spatial_plots}
    %\vspace{-2mm}
  \end{figure}

\begin{figure}[t!] % [t!] % [htbp]
  \centering
  \includegraphics[width=3.45in]{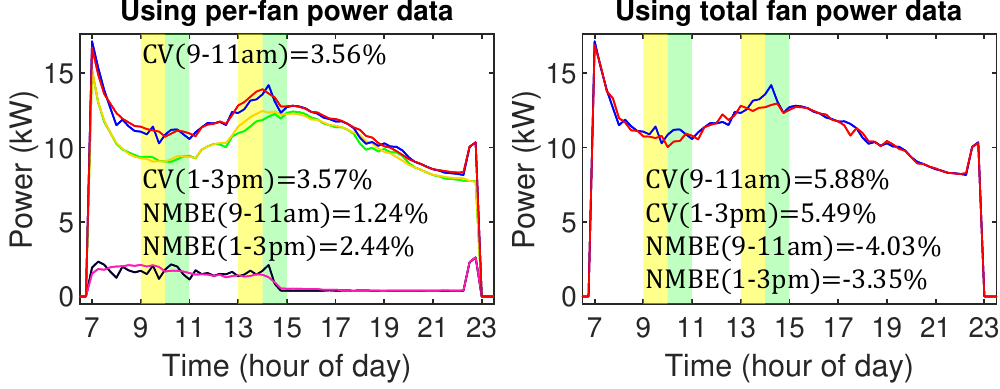}\\
  \includegraphics[width=3.2in]{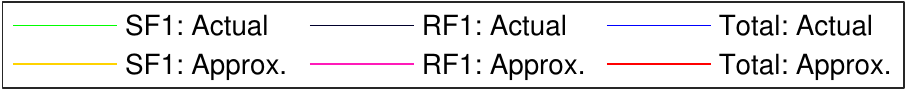}
  \caption{Actual and approximated per-fan and$/$or total fan power traces based on
  per-fan power data and total fan power data (BBB, Aug. 14, 2017).}
  \label{per_fan_sum_all_example}
%\vspace{-2mm}
\end{figure}

Here, we apply the tensor completion method to total fan power data
obtained by summing the original per-fan data across fans.
Fig.~\ref{spatial_plots} shows the error metrics, based on 15-minute interval data, obtained using per-fan data versus total fan power data. The tensor completion method with per-fan power data
outperforms
the degenerate\footnote{By ``degenerate'',
{\color{black}we} mean that the tensor completion method is
applied to the two-way $T \times S$ tensor of total fan power data here, while the
method is originally developed for the three-way $T \times N \times S$ tensor of per-fan power data.}
version of the method with total fan power data.
This result supports the use of our 3-dimensional per-fan power data based tensor completion
method, i.e., baselining total fan power according to
dominant per-fan power patterns that are consistent among different fans and
over different days.

Fig.~\ref{per_fan_sum_all_example}
shows an example of the per-fan and total fan power traces approximated by the tensor completion method with per-fan data, and the total fan power trace
approximated by the degenerate version of the method with total fan power data.
As shown in Fig.~\ref{per_fan_sum_all_example} (Left), the proposed
method approximates power traces of supply fan 1 and return fan 1,
which are summed to approximate the total fan power.
During event windows where the approximation acts as the estimated baseline,
the per-fan estimations are smooth. The
positive and negative errors somewhat balance each other out over time,
and the estimation errors of two fans generally balance each other out.
The resulting total fan power baseline has lower CV and NMBE.
The other case in Fig.~\ref{per_fan_sum_all_example} (Right) that directly
approximates the total fan power trace produces a less smooth baseline estimate and also
higher CV and NMBE.

\subsection{Huber Loss Function versus $L_2$-Norm Loss Function}

\begin{figure} \centering
%   \begin{subfigure}{.2\linewidth} \centering
    \includegraphics[scale=0.62]{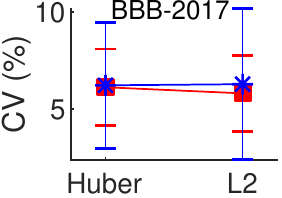}%
%     \label{cv1_huber}
%   \end{subfigure}\hfil
%   \begin{subfigure}{.2\linewidth} \centering
    \includegraphics[scale=0.62]{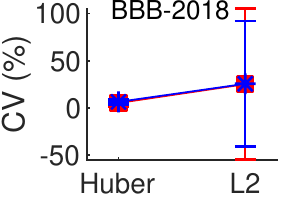}%
%     \label{cv2_huber}
%   \end{subfigure}\hfil
%   \begin{subfigure}{.2\linewidth} \centering
    \includegraphics[scale=0.62]{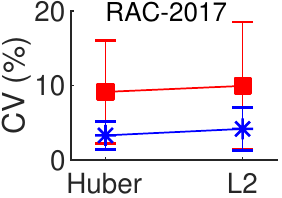}%
%     \label{cv3_huber}
%   \end{subfigure}\hfil
%   \begin{subfigure}{.2\linewidth} \centering
    \includegraphics[scale=0.62]{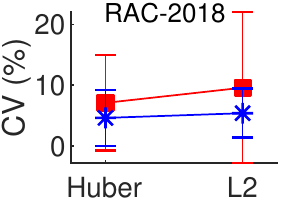}%
%     \label{cv4_huber}
%   \end{subfigure}\hfil
%   \begin{subfigure}{.2\linewidth} \centering
    \includegraphics[scale=0.62]{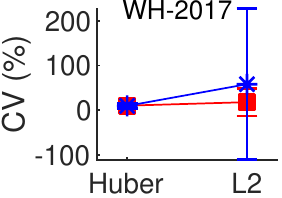}\\[1.5mm]
%     \label{cv5_huber}
%   \end{subfigure}\\
%   \vspace{-2.5mm}
%   \begin{subfigure}{.2\linewidth} \centering
    \includegraphics[scale=0.62]{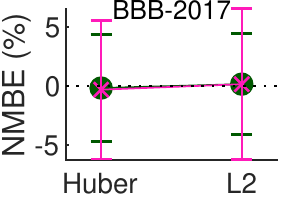}%
%     \label{MBE1_huber}
%   \end{subfigure}\hfil
%   \begin{subfigure}{.2\linewidth} \centering
    \includegraphics[scale=0.62]{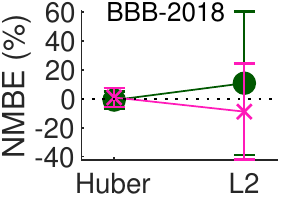}%
%     \label{MBE2_huber}
%   \end{subfigure}\hfil
%   \begin{subfigure}{.2\linewidth} \centering
    \includegraphics[scale=0.62]{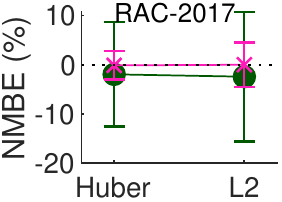}%
%     \label{MBE3_huber}
%   \end{subfigure}\hfil
%   \begin{subfigure}{.2\linewidth} \centering
    \includegraphics[scale=0.62]{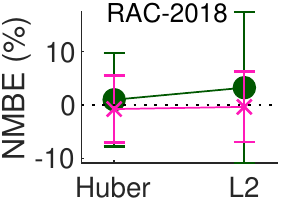}%
%     \label{MBE4_huber}
%   \end{subfigure}\hfil
%   \begin{subfigure}{.2\linewidth} \centering
    \includegraphics[scale=0.62]{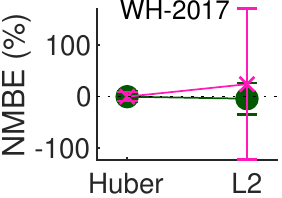}\\
%     \label{MBE5_huber}
%   \end{subfigure}\\
%     \vspace{-3.75mm}
    \includegraphics[scale=0.74]{spatial_legend-crop}
    \caption{Mean values of CV and NMBE of the tensor completion method with the Huber and
    $L_2$-norm loss functions. (The error bars represent standard deviations.)}
    \label{compare_huber_L2}
    %\vspace{-2mm}
  \end{figure}

  \begin{figure} \centering
  %   \begin{subfigure}{.25\linewidth} \centering
      \includegraphics[scale=0.75]{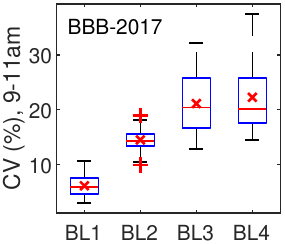}\hspace{5pt}
  %   \end{subfigure}\hfil
  %   \begin{subfigure}{.25\linewidth} \centering
      \includegraphics[scale=0.75]{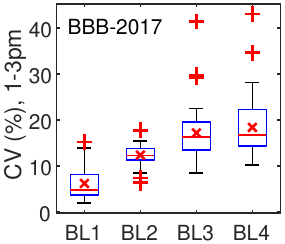}\hspace{5pt}
  %   \end{subfigure}\hfil
  %   \begin{subfigure}{.25\linewidth} \centering
      \includegraphics[scale=0.75]{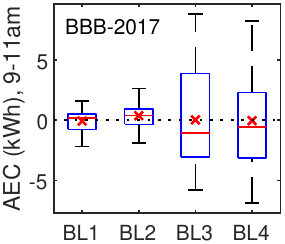}\hspace{5pt}
  %   \end{subfigure}\hfil
  %   \begin{subfigure}{.25\linewidth} \centering
      \includegraphics[scale=0.75]{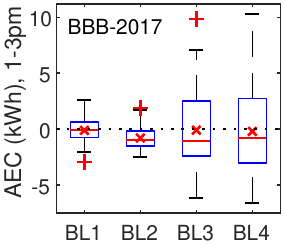}\\[2mm]
  %   \end{subfigure}\\[1mm]
  %   \begin{subfigure}{.25\linewidth} \centering
      \includegraphics[scale=0.75]{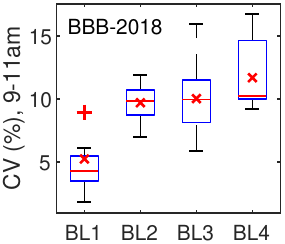}\hspace{5pt}
  %   \end{subfigure}\hfil
  %   \begin{subfigure}{.25\linewidth} \centering
      \includegraphics[scale=0.75]{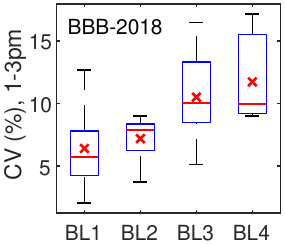}\hspace{5pt}
  %   \end{subfigure}\hfil
  %   \begin{subfigure}{.25\linewidth} \centering
      \includegraphics[scale=0.75]{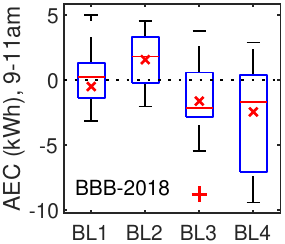}\hspace{5pt}
  %   \end{subfigure}\hfil
  %   \begin{subfigure}{.25\linewidth} \centering
      \includegraphics[scale=0.75]{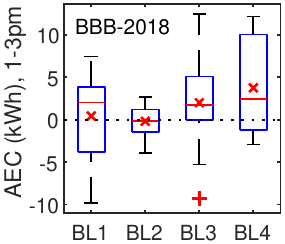}\\[2mm]
  %   \end{subfigure}\\[1mm]
  %   \begin{subfigure}{.25\linewidth} \centering
      \includegraphics[scale=0.75]{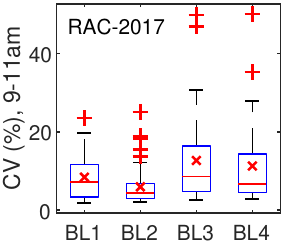}\hspace{5pt}
  %   \end{subfigure}\hfil
  %   \begin{subfigure}{.25\linewidth} \centering
      \includegraphics[scale=0.75]{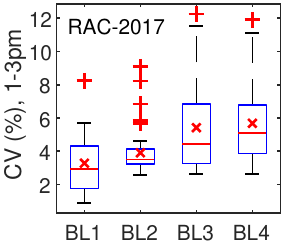}\hspace{5pt}
  %   \end{subfigure}\hfil
  %   \begin{subfigure}{.25\linewidth} \centering
      \includegraphics[scale=0.75]{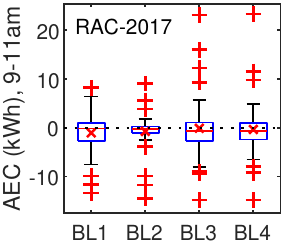}\hspace{5pt}
  %   \end{subfigure}\hfil
  %   \begin{subfigure}{.25\linewidth} \centering
      \includegraphics[scale=0.75]{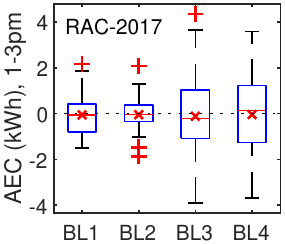}\\[2mm]
  %   \end{subfigure}\\[1mm]
  %   \begin{subfigure}{.25\linewidth} \centering
      \includegraphics[scale=0.75]{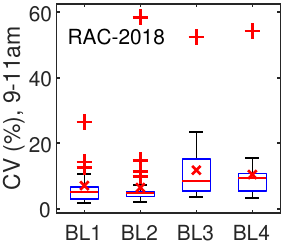}\hspace{5pt}
  %   \end{subfigure}\hfil
  %   \begin{subfigure}{.25\linewidth} \centering
      \includegraphics[scale=0.75]{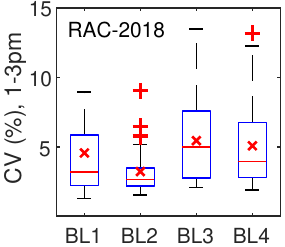}\hspace{5pt}
  %   \end{subfigure}\hfil
  %   \begin{subfigure}{.25\linewidth} \centering
      \includegraphics[scale=0.75]{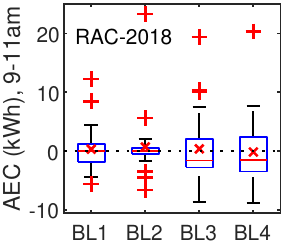}\hspace{5pt}
  %   \end{subfigure}\hfil
  %   \begin{subfigure}{.25\linewidth} \centering
      \includegraphics[scale=0.75]{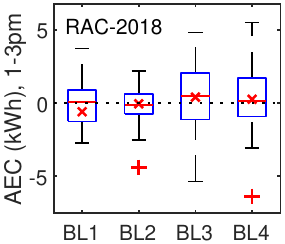}\\[2mm]
  %   \end{subfigure}\\[1mm]
  %   \begin{subfigure}{.25\linewidth} \centering
      \includegraphics[scale=0.75]{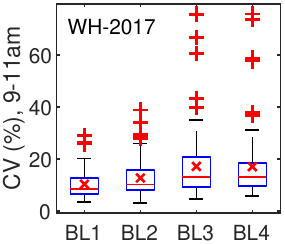}\hspace{5pt}
  %   \end{subfigure}\hfil
  %   \begin{subfigure}{.25\linewidth} \centering
      \includegraphics[scale=0.75]{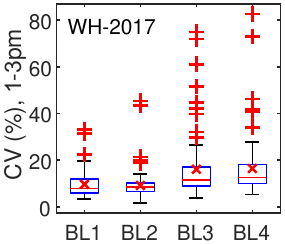}\hspace{5pt}
  %   \end{subfigure}\hfil
  %   \begin{subfigure}{.25\linewidth} \centering
      \includegraphics[scale=0.75]{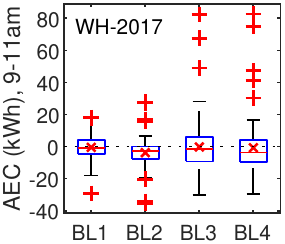}\hspace{5pt}
  %   \end{subfigure}\hfil
  %   \begin{subfigure}{.25\linewidth} \centering
      \includegraphics[scale=0.75]{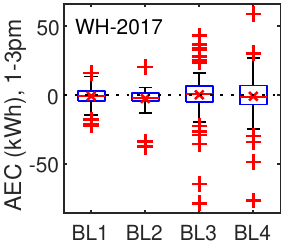}
  %   \end{subfigure}
      \caption{Boxplots of CV and AEC for different building-years.
      (BL1: Tensor; BL2: Linear interpolation; BL3: 5-day average; BL4: Nearest3of6.)}
      \label{boxplots_compare}
      % \vspace{-5mm}
  %\vspace{-3mm}
  \end{figure}

Here, we test the $L_2$-norm loss function and compare it to the Huber loss function, used to generate the previous results.
Fig.~\ref{compare_huber_L2} shows the comparison using 15-min interval per-fan power data.
For building-years BBB-2017, RAC-2017 and RAC-2018,
the two loss functions
produce comparable results, with the Huber loss function producing slightly better results
in general.
However, for BBB-2018 and WH-2017, the $L_2$-norm loss function produces
results with much larger baseline errors.
BBB-2018 is the smallest data set, with only 16
baseline days. It has an obvious outlier on one baseline day when 6 fans suddenly stop working for about 30 minutes.
WH-2017 is the largest data set, but with more visible outliers than other building-years.
For example, its day mode HVAC operational period is normally 5am-5pm or 4am-9pm;
however, on several days the period is different from the normal ones and also different from the other abnormal ones.
These results imply higher robustness to outliers with the Huber loss function compared to the $L_2$-norm loss function.

\subsection{Comparison with Other Baseline Methods}

\begin{figure*}[t!] % [t!] % [htbp]
    \centering
    \includegraphics[width=6.9in]{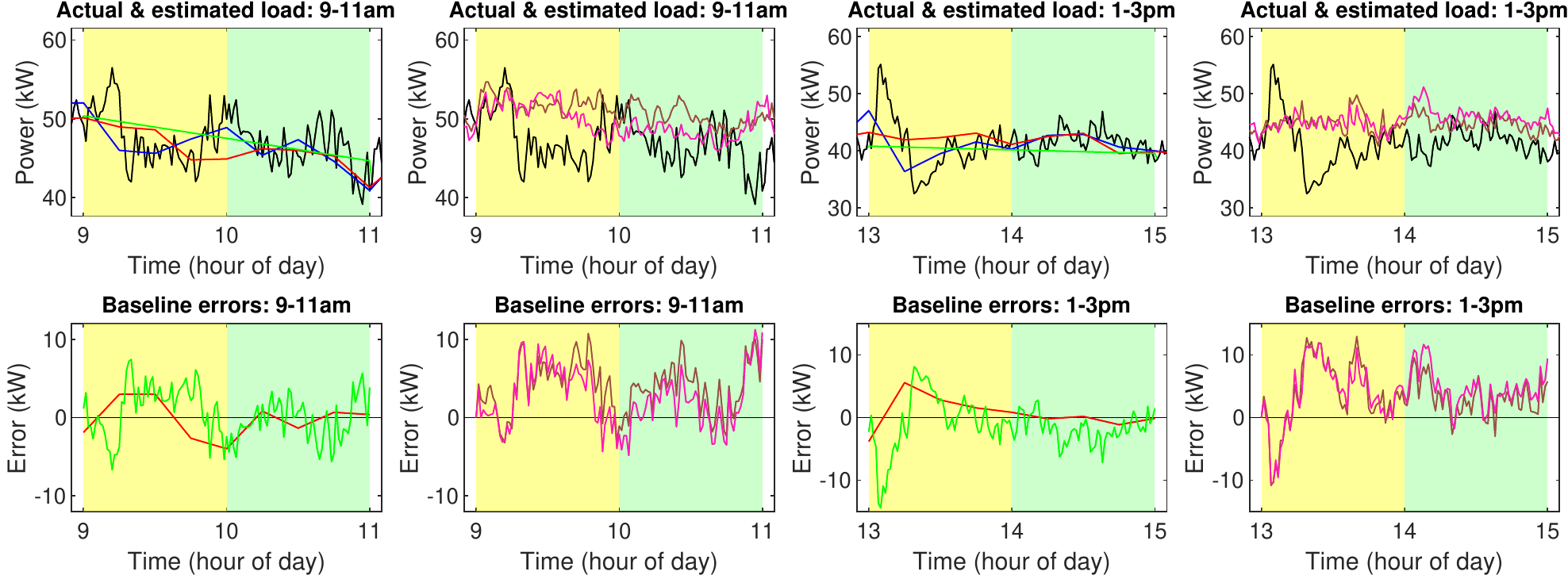}\\
    \includegraphics[width=3.45in]{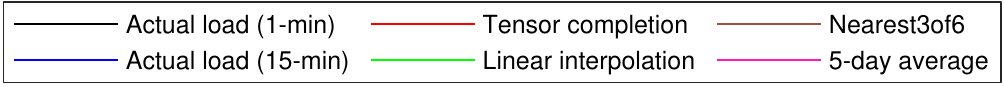}
    \caption{Actual and estimated total fan power baselines and the estimation errors
    (WH, Jul. 19, 2017).
    CV(9-11am) of the tensor completion method$/$linear interpolation method$/$5-day average method$/$nearest3of6 method are:
    5.50\%$/$6.66\%$/$10.93\%$/$9.65\%;
    CV(1-3pm): 6.95\%$/$9.68\%$/$13.69\%$/$14.59\%;
    AEC(9-11am): -0.601kWh$/$0.832kWh$/$7.974kWh$/$5.090kWh;
    AEC(1-3pm): 1.449kWh$/$-2.392kWh$/$7.088kWh$/$8.437kWh.}
    \label{best3_vs_TD_example}
%\vspace*{-3mm}
\end{figure*}

\begin{table}[t!]
    \centering
    \caption{AEC of the Tensor Completion Method and Linear Interpolation Method}
    \begin{tabular}{c|c|cc}
      \hline
      \multicolumn{2}{c}{}  & Tensor completion & Linear interpolation \\
      \hline
      \multirow{5}[0]{*}{\scriptsize\rotatebox{90}{\tabincell{c}{AEC (kWh), \\9-11am: \\Bias$\pm$95\%CI}}}
      & BBB-2017  &  \textbf{-0.056 $\pm$ 0.243}     &  0.390  $\pm$ 0.273 \\
      %\cline{2-4}
      & BBB-2018 & \textbf{-0.529} $\pm$ 1.732  & 1.543 $\pm$ 1.007 \\
      %\cline{2-4}
      & RAC-2017 & -1.034 $\pm$ 1.368 & \textbf{-0.728 $\pm$ 0.885} \\
      %\cline{2-4}
      & RAC-2018 & \textbf{0.288 $\pm$ 1.387} & 0.664 $\pm$ 1.624 \\
      %\cline{2-4}
      & WH-2017 & \textbf{-0.495 $\pm$ 1.685} & -3.815 $\pm$ 2.016\\
      \hline
      \multirow{5}[0]{*}{\scriptsize\rotatebox{90}{\tabincell{c}{AEC (kWh), \\1-3pm: \\Bias$\pm$95\%CI}}}
      & BBB-2017  & \textbf{-0.105} $\pm$ 0.361 & -0.786 $\pm$ 0.261   \\
      %\cline{2-4}
      & BBB-2018 & 0.433 $\pm$ 2.427  & \textbf{-0.192 $\pm$ 0.928} \\
      %\cline{2-4}
      & RAC-2017 & -0.059 $\pm$ 0.243 & \textbf{-0.051 $\pm$ 0.194} \\
      %\cline{2-4}
      & RAC-2018 & -0.603 $\pm$ 1.380 & \textbf{-0.060 $\pm$ 0.474} \\
      %\cline{2-4}
      & WH-2017 & \textbf{-0.729} $\pm$ 1.532 & -2.510 $\pm$ 1.349 \\
      \hline
  \end{tabular}%
  \label{AEC_table}%
%\vspace{-3mm}
\end{table}%
  % \begin{table}[t!]
  %   \centering
  %   \caption{...}
  %   % \setlength{\tabcolsep}{1.3pt}
  %   \begin{tabular}{|c|c|c|c|}
  %     \hline
  %     \multicolumn{2}{|c|}{-----------}  & Tensor completion & Linear interpolation \\
  %     \hline
  %     \multirow{5}[0]{*}{\scriptsize\rotatebox{90}{\tabincell{c}{AEC (kWh), \\9-11am: \\Bias$\pm$95\%CI}}}
  %     & BBB-2017  &  \cellcolor{green!70}-0.056 $\pm$ 0.243     &  0.390  $\pm$ 0.273 \\
  %     \cline{2-4}
  %     & BBB-2018 & \cellcolor{green!20}-0.529 $\pm$ 1.732  & \textbf{\textit{1.543}} $\pm$ 1.007 \\
  %     \cline{2-4}
  %     & RAC-2017 & \textbf{\textit{-1.034}} $\pm$ 1.368 & \cellcolor{green!70}-0.728 $\pm$ 0.885 \\
  %     \cline{2-4}
  %     & RAC-2018 & \cellcolor{green!70}0.288 $\pm$ 1.387 & 0.664 $\pm$ 1.624 \\
  %     \cline{2-4}
  %     & WH-2017 & \cellcolor{green!70}-0.495 $\pm$ 1.685 & \textbf{\textit{-3.815}} $\pm$ 2.016\\
  %     \hline
  %     \multirow{5}[0]{*}{\scriptsize\rotatebox{90}{\tabincell{c}{AEC (kWh), \\1-3pm: \\Bias$\pm$95\%CI}}}
  %     & BBB-2017  & \cellcolor{green!20}-0.105 $\pm$ 0.361 & -0.786 $\pm$ 0.261   \\
  %     \cline{2-4}
  %     & BBB-2018 & 0.433 $\pm$ 2.427  & \cellcolor{green!70}-0.192 $\pm$ 0.928 \\
  %     \cline{2-4}
  %     & RAC-2017 & -0.059 $\pm$ 0.243 & \cellcolor{green!70}-0.051 $\pm$ 0.194 \\
  %     \cline{2-4}
  %     & RAC-2018 & -0.603 $\pm$ 1.380 & \cellcolor{green!70}-0.060 $\pm$ 0.474 \\
  %     \cline{2-4}
  %     & WH-2017 & \cellcolor{green!20}-0.729 $\pm$ 1.532 & \textbf{\textit{-2.510}} $\pm$ 1.349 \\
  %     \hline
  % \end{tabular}%
  % \label{AEC_table}%
  % \end{table}%

Next, we compare the tensor completion method (using 15-minute interval per-fan power data and the Huber loss function) with three other baseline methods:
1)~\emph{Linear interpolation}~\cite{Beil2015}~\cite{Keskar2018}:
This method uses least squares to fit a linear baseline to the total fan power data over the 5-minute periods just before
and immediately after the event window.
2)~\emph{5-day average}~\cite{Coughlin2009}:
The average load over the 5 most recent baseline days is used to estimate the baseline.
3)~\emph{Nearest3of6}~\cite{SLei2019}:
Among the 6 most recent baseline days, the 3 days with electricity consumption outside of
event windows that is closest to that on the event day are averaged to estimate the baseline.
{\color{black}Following \cite{SLei2019},} the three benchmark methods use 1-minute interval total fan power data
here. {\color{black}The impact of data granularity on the benchmark methods' performance will be explored
in future work.} A simple additive adjustment,
which vertically shifts the baseline estimate so that it is equal to the actual load
just prior to the event window, is also applied to the latter two methods~\cite{SLei2019}.

Fig.~\ref{boxplots_compare} shows the comparison.
The tensor completion method and linear interpolation method have better performance
than the 5-day average and Nearest3of6 methods.
The tensor completion method generally outperforms the linear interpolation method for
BBB-2017, BBB-2018 and WH-2017.
For RAC-2017 and RAC-2018
their accuracy is similar.
Fig.~\ref{best3_vs_TD_example}
shows an example of the actual and estimated total fan power baselines.
As can be seen, the two averaging methods do not estimate the
minute-scale variation of the actual baseline very well, producing
larger errors.
The linear interpolation method does not capture the variation at all.
Instead, it expects positive and negative errors to balance out over time, which is appropriate for some DR applications that only need the average deviation over the event.
In this example, it performs well; however, the tensor completion method outperforms it.

Table~\ref{AEC_table} reports mean values and 95\% confidence intervals for the AEC
for the tensor completion method and~linear~interpolation method.
{\color{black}The 95\% confidence intervals are calculated by
$\david{\operatorname{mean}}(AEC) \pm 1.96 \times \david{\operatorname{std}}(AEC)/ \sqrt{S}$, where $S$ denotes
the number of baseline days in a data set that the baseline method is tested on.}
Note that
the bias of AEC is especially important for DR applications
such as financial settlement.
Table~\ref{AEC_table} indicates that the tensor completion method generally outperforms
the linear interpolation method in baselining fan power within the
morning event window, while the interpolation method is generally better within the afternoon event window.
Still, the tensor completion method has a more consistent performance,
while the interpolation method ends up with a much larger bias in BBB-2018 (for the morning event window) and WH-2017 (for both event windows).

\section{Conclusions and Future Work}\label{conclusion_future_work}

% !TEX root = main.tex

This paper investigated the use of tensor completion
for baseline estimation of commercial building HVAC system fan power.
The proposed method estimates baseline fan power
by forming a tensor from submetered per-fan power data and
then imputing the counterfactual measurements from DR event windows
to encourage correlations along the dimensions of time, fan, and day.
These estimates of counterfactual fan power
are taken from a low-rank GCP tensor decomposition
that approximates the fan power data tensor on measurements of baseline days and
of the event day but outside DR event window(s).
We compared the proposed method with the linear interpolation,
5-day average, and Nearest3of6 methods
using data collected at the University of Michigan
and found that the proposed method generally has the best performance.

An important direction for future work is
developing methods for selecting the tensor rank.
Various methods might be considered,
e.g., one could select the rank that best estimates the
measurements on a validation set of the data,
but more work is needed to assess and understand the trade-off
between data overfit and approximation error
discussed in \cref{sec:proposal:rank}.
These tradeoffs can also be influenced if we incorporate regularization
in the GCP tensor decomposition objective \cref{eq:gcp:objective}
as described in~\cite[Section 4.2]{hong:2018:gcp}.
For example, regularizing the temporal factors of $\cM$ to be piecewise smooth
may help mitigate data overfit, allowing for decompositions of higher ranks.

Another avenue for future work is to consider other tensor completion approaches,
such as those described in the recent survey \cite{song2019tca}.
In particular, many approaches exploit low-rank structure
with respect the Tucker decomposition \cite[Section~4]{kolda:2009:tda},
e.g., \cite{zhang2019cel} is a recent proposal based on Tucker rank.
Exploring these tensor completion approaches and their tradeoffs
is an important and promising direction.

{\color{black}
\section*{Acknowledgement}
The authors thank Rishee K. Jain, Jeremiah X. Johnson and
Aditya Keskar for valuable discussions.}
% The authors also thank the reviewers for their
% insightful comments and constructive suggestions.}

% The 'triggered' command can be changed if desired:


%\bibliographystyle{IEEEtran}
% \bibitem{Shell}
% M.~Shell, \emph{How to Use the IEEEtran Latex Class}, Latex Archive Contents, \verb+http://www.ieee.org/conferences_events/+ \verb+conferences/publishing/templates.htm+

% \end{thebibliography}

% \vspace{-1.25pt}
\bibliographystyle{IEEEtran}
% \vspace{-1.75pt}
\bibliography{IEEEabrv,references}

\end{document}